# Applying the Estimand and Target Trial frameworks to external control analyses using observational data: a case study in the solid tumor setting

Letizia POLITO[1]*, Qixing LIANG[2]*, Navdeep PAL[3], Philani MPOFU[2], Ahmed SAWAS[2], Olivier HUMBLET[2], Kaspar RUFIBACH[1], Dominik HEINZMANN[4]

[1]F. Hoffmann-La Roche Ltd, Switzerland

[2]Flatiron Health, New York, NY, USA

[3]Genentech, Inc., South San Francisco, CA, USA

[4]Novo Nordisk, Denmark

*co-first authors

**Corresponding author:**

Letizia Polito, PhD

Product Development Data Science, F. Hoffmann–La Roche,

Grenzacherstrasse 124, Basel CH-4070, Switzerland

letizia.polito@roche.com

+41799004136

**Conflict of Interests:** At the time of this study, QL, PM, AS and OH report employment at Flatiron Health Inc., an independent subsidiary of the Roche Group, and stock in Roche. At the time of this study, LP, KR, and DH report employment at F. Hoffmann-La Roche and stock ownership in Roche. At the time of this study, NP reports employment at Genentech Inc, and stock ownership at Roche/Genentech.

**Study Funding:** This study was sponsored by Flatiron Health, Inc., which is an independent subsidiary of the Roche Group.

# Abstract

In causal inference, the correct formulation of the scientific question of interest is a crucial step. Here we apply the estimand framework to a comparison of the outcomes of patient-level clinical trials and observational data to help structure the clinical question. In addition, we complement the estimand framework with the target trial framework to address specific issues in defining the estimand attributes using observational data and discuss synergies and differences of the two frameworks. Whereas the estimand framework proves useful to address the challenge that in clinical trials and routine clinical practice patients may switch to subsequent systemic therapies after the initially assigned systematic treatment, the target trial framework supports addressing challenges around baseline confounding and the index date.

We apply the combined framework to compare long-term outcomes of a pooled set of three previously reported randomized phase 3 trials studying patients with metastatic non-small cell lung cancer receiving front-line chemotherapy (randomized clinical trial cohort) and similar patients treated with front-line chemotherapy as part of routine clinical care (observational comparative cohort). We illustrate the process to define the estimand attributes and select the estimator to estimate the estimand of interest while accounting for key baseline confounders, index date, and receipt of subsequent therapies. The proposed combined framework provides more clarity on the causal contrast of interest and the estimator to adopt and thus facilitates design and interpretation of the analyses.

# Introduction

Several causal inference frameworks including the estimand framework (EF), target trial emulation framework (TTF) and PICO frameworks exist to help define a precise scientific question for comparative assessments in clinical research and development (i.e., whenever a treatment effect is estimated or a hypothesis related to a treatment effect is tested)[1]. There are overlapping but complementary elements in these frameworks, suggesting the potential for a combined application. However, this presents challenges to investigators as there are limited practical examples and guidance for combined application of the frameworks.

The EF is one such framework which has increasingly been adopted by health authorities and pharmaceutical companies since initial publication in August 2017.[2] The EF enables specifying a precise scientific question by using five attributes that define the estimand (i.e., the treatment effect of interest or the "what to estimate"). These five interrelated attributes are population, treatment, variable of interest (endpoint), intercurrent event handling, and the summary measure. An intercurrent event is an event occurring after treatment initiation that affects either the interpretation or the existence of the measurements associated with the endpoint. For example, if performing a comparative assessment on overall survival between two different treatments, candidates for intercurrent events include, among others, early discontinuation of treatment or treatment switching after disease progression. In general, the definition of the estimand comes first and is derived from the scientific objective of the trial or study. Together with considerations about missing data, the framework then informs the choice of the estimator. The addendum acknowledges that usually an iterative process will be necessary to reach an estimand that is of clinical relevance for decision making *and* for which a reliable estimate can be computed. If it turns out to not be possible to develop an appropriate trial design or to derive an adequately reliable estimate for a particular estimand, an alternative estimand, trial design, and method of analysis may need to be considered. However, practical examples in the literature describing such an iterative process to redefine an initial target estimand and also considering aspects of identifiability (and hence the estimator) are limited. While the focus of the ICH E9 addendum is on randomized clinical trials (RCTs), the principles are also applicable whenever estimating a treatment effect (i.e., also for single arm trials and observational studies). However, estimation of a causal effect from observational data often has additional challenges compared to an RCT. These include bias due to baseline confounding, selection bias, missing data and defining the index date for comparison.

The TTF is another causal framework that can be used to more precisely specify the scientific question in a comparative assessment.[3] TTF complements the EF by addressing gaps related to the analysis of observational data and is the application of design principles of an RCT to the specific setting of a non-randomized comparative assessment.[3-6] TTF entails defining a hypothetical randomized trial to address a precise scientific question, and then further specifying how it can be emulated (i.e., approximated) by non-randomized data. The essential components of a target trial protocol are eligibility criteria, treatment strategies, treatment assignment, start and end of follow-up, outcomes, causal contrasts, and the analysis approach (estimator).[3] The framework can also be utilized when a combination of clinical trial and observational data is used, for example to contextualize a single arm clinical trial with observational data.[7] Combining the EF and the TTF provides a structured approach to enhance the scientific rigor for causal inference for observational and/or non-randomized data. Together they bring more transparency on the causal estimand, supporting specifying the attributes of the estimand and the assumptions to be made to draw causal conclusions.

Another framework that aims to define the precise scientific question include the PICO framework,[8] that has traditionally been used in epidemiology for observational studies. The EF and TTF both extend the PICO framework, with the former adding intercurrent events and the population-level summary measure, and the latter adding the causal contrast, assignment procedures, and the start/end of follow-up.

In this study we jointly apply the EF and TTF to perform a comparative effectiveness assessment in patients with non-small cell lung cancer (NSCLC) using data from a set of pooled control arms of three RCTs[(9-11)] as well as electronic health record (EHR)-derived de-identified observational data. The objective of our case study was to determine whether there is a difference in overall survival (OS) between patients with metastatic NSCLC receiving front-line chemotherapy in pivotal trials vs patients with metastatic NSCLC who received front-line chemotherapy as part of routine care, had patients not received a subsequent therapy. This case study contributes to the important goal in pharmacoepidemiology of assessing whether observational data (including EHR-derived) can emulate (and thus supplement or replace, e.g. for regulatory decision-making) the control arm of a RCT. The iterative process (as indicated in the EF) to arrive at the final scientific question is illustrated in the *Methods* section. The case study illustrates the applicability of the EF to observational data, and furthermore complements the EF with the TTF to account for specific challenges in observational data not directly addressed by the EF (and vice versa, as the handling of intercurrent events is not explicitly addressed in the TTF). In sum, the present study provides insights into where the two frameworks are complementary, and provides a practical example of jointly applying them.

# Methods

## Applying the frameworks to the research question

Before discussing details on the joint application of the EF and TTF to the final scientific question, we want to provide insights into the iterations to arrive at the final question as we hope this would provide practical guidance on the iterative process outlined in the EF. We were interested in comparing the treatment effect of the same front-line treatment given in a clinical trial and in clinical practice when subsequent treatments would be similar. We started with the scientific question: “Is there a difference in overall survival (OS) between patients with metastatic NSCLC receiving front-line chemotherapy in pivotal trials vs patients with metastatic NSCLC who received front-line chemotherapy as part of routine care?”. EHR-derived observational data from routine clinical practice suggest a larger heterogeneity in subsequent second-line cancer treatments as compared to a clinical trial setting.[12] This difference in the range of potential subsequent therapies may introduce complexities in estimating causal treatment effects for longer-term outcomes such as OS and ultimately complicate interpretation. As a consequence, the initial research question has been iterated to: “Is there a difference in overall survival (OS) between patients with metastatic NSCLC receiving front-line chemotherapy in pivotal trials vs patients with metastatic NSCLC who received front-line chemotherapy as part of routine care, *had patients not received a subsequent therapy*?”. Hence, instead of considering the entire treatment strategy (front-line and subsequent therapy) which is complicated by heterogeneity in subsequent therapies among clinical trial and clinical practice settings, the iteration resulted in the scientific question of treatment effect of the front-line regimes.

Now we focus on the joint application of the EF and TTF to the final scientific question. Table 1 displays the EF/TTF attributes that define the estimand aligned with the scientific research question. We define the hypothetical target trial structured according to the estimand framework, and we define the study that attempts to emulate it leveraging elements from the EF and TTF. The average treatment effect on the treated (ATT) is the estimand of primary interest. This is the treatment effect difference of using front-line chemotherapy in a clinical trial versus in clinical practice, and hence the target population is defined by the clinical trials population.

## Data source

### Clinical trial data

Individual patient-level data (PLD) were used from Roche-sponsored phase III, open-label randomized clinical trials IMpower130 (ClinicalTrials.gov identifier: NCT02367781), 131 (ClinicalTrials.gov identifier: NCT02367794) and 132 (ClinicalTrials.gov identifier: NCT02657434). Methods and primary findings have been previously reported.[9-11] These three trials included patients who were chemotherapy-naive and had stage IV NSCLC. OS was the primary endpoint for the three trials. To address the objective of the present study only the PLD from the control arms were used. The control arms received platinum-based chemotherapy as follows:

- IMpower130 included patients with non-squamous NSCLC treated with carboplatin plus nab-paclitaxel
- IMpower131 included patients with squamous NSCLC treated with carboplatin plus nab-paclitaxel
- IMpower132 included patients with non-squamous NSCLC treated with carboplatin or cisplatin plus pemetrexed

As these three clinical trial control arms had similar settings in terms of disease, therapy, and inclusion/exclusion criteria and also had similar survival outcomes such as median survival time (Appendix Figure 1), they were pooled together to increase the sample size and are collectively referred to as the RCT arm in this study.

### Observational data

The observational comparator (OC) arm of this study was developed using the nationwide Flatiron Health EHR-derived de-identified database. This longitudinal database is comprised of patient-level structured data (e.g., laboratory values and prescribed treatments) and unstructured data (e.g., biomarker reports) curated from technology-enabled chart abstraction from physicians' notes and other documents. During the study period, the de-identified data originated from approximately 280 cancer clinics (approximately 800 sites of care, primarily community-based cancer centers).[13,14] Institutional Review Board approval of the study protocol was obtained prior to study conduct, and included a waiver of informed consent.

## Cohort selection/Study sample

The OC cohort was selected to align to the clinical trials eligibility (inclusion/exclusion) criteria of the three trials. To be eligible for entry into the Flatiron Health Research Database, the patient's record must include >1 visit to a community oncology clinic documented in the EHR, and have confirmation of advanced NSCLC diagnosis and histological subtype (squamous vs. non-squamous histology) through a review of unstructured data (ie, clinical notes, radiology reports, or pathology reports). A start date of front-line therapy for advanced or metastatic NSCLC on or after April 16, 2015 and on or before May 31, 2017 to match the clinical trials start and end dates of enrollment was also required. Patients with an ECOG performance status (PS) of 0, 1, or unknown were included. Patients had to have received at least one administration of regimens of interest (i.e., carboplatin plus paclitaxel/nab-paclitaxel, carboplatin or cisplatin plus pemetrexed). Patients who had potentially incomplete historical treatment data (i.e., >90-day gap between advanced diagnosis and structured activity in the EHR), therapy within 6 month prior to start of front-line therapy for advanced stage disease, receipt of clinical study drug or multiple primary tumors were excluded. Patients with missing information or known to have a sensitizing mutation in the EGFR gene or an ALK fusion oncogene were excluded. All patients were followed until July 18th 2019. Detailed inclusion/exclusion criteria were included in Appendix Table 5.

## Statistical analyses

The following estimation approach to target the ATT estimand with attributes as specified in Table 1 has been applied. First, the inverse probability of treatment weighting (IPTW) method was used to balance baseline patient characteristics between the trial arm and the external control arm. A multiple logistic model was used to estimate propensity scores that are defined as probabilities of being assigned to the RCT group conditional on all confounders (Appendix Table 4) that were selected based on clinical experts' knowledge and availability of the relevant variables. Given that we target the ATT as outlined above, patients from clinical trials were given a weight of one while weights for OC patients were defined as the ratio of the estimated propensity score (PS) to one minus the estimated PS (i.e., odds of being treated in the clinical setting). After IPTW weighting, differences in baseline characteristics were assessed through standardized mean differences (SMD) (Table 2). Patient characteristics were considered balanced if SMD < 0.10.[(15)] The weighted population was used in the subsequent analyses.

Secondly, we artificially censored patients at the time of receipt of first second-line treatment and used inverse probability of censoring weighting (IPCW) method to estimate weights for the follow-up information for the remaining patients using both baseline and time-varying variables which are likely to impact treatment switching based on clinical experts' knowledge to adjust for any potential confounding created by the artificial censoring. Specifically, we fit a Cox model for each group that was used to estimate the probability of not being censored by time $t$ given baseline and time-varying covariates listed in Table 7 for the specific group. The IPCW weights are calculated as the inverse of the conditional probability of not being censored. We truncated the follow-up time at Month 21 because there were few patients remaining in the RCT group after Month 21 and thus the positivity assumption was unlikely to hold. Then, in order to reduce variance of the weighted estimator, we calculated the stabilized IPCW weight[(15)] which is the probability of not being censored conditional on selected baseline covariates divided by the probability of not being censored conditional on both baseline and time-varying covariates. Standardized weighted mean and proportion difference (SMD) was used to assess differences in censoring confounders between the two groups after weighting using the same definition of balance as above.

The treatment effects were estimated using weighted survival analysis methods. Specifically, we estimated the hazard ratio (HR) using a IPTW-IPCW weighted Cox proportional hazard model and the 95% CI for the HR using bootstrap approach.[16] We also used the IPTW-IPCW weighted Kaplan–Meier method to compute OS survival function estimates and weighted log-rank test to compare across groups. Hence the double weighting estimation approach targets the ATT estimand with attributes of the EF & TTF framework as specified in Table 1.

Missing values for covariates with missing rate less than 30% were imputed using median (for age and time from initial diagnosis to index date) or mode (smoking history). Covariates with more than 30% of values missing (i.e., ECOG PS) were excluded from the analysis. We performed a sensitivity analysis by analyzing the whole follow-up time periods for RCT and OC groups instead of truncating them at Month 21. Also, to evaluate to what extent our estimation methods remove the potential bias on OS due to baseline confounders and intercurrent events, we performed the traditional IPTW-only method that adjusts for baseline characteristics but not intercurrent events in terms of K-M estimate and HR and compared to our proposed method. To follow the structure of the Estimand framework, we consider this IPTW-only estimation as a supplementary analysis because it estimates an estimand different from our target estimand.

R (3.6.1) was used for the analyses.

# Results

## Cohort characteristics

A total of 849 patients were in the RCT arm and 3,340 patients were in the OC arm (refer to Appendix Table 2 for the OC cohort attrition table). Demographic and clinical characteristics of the study sample at baseline are presented in Table 2 (and in Appendix Table 3 stratified by RCT). Significant differences between the RCT and OC arms were observed in age, gender, race, ECOG-PS,  tumor diagnosis type (de novo Stage IV/recurrent disease), histology, time from initial diagnosis to index date, and treatment type. Patients in the OC arm were older, with higher percentage of females, races other than White and Asian, recurrent disease and non-squamous histology, with shorter time from initial diagnosis to index date, and less frequently treated with carboplatin plus paclitaxel/nab-paclitaxel.

The percentage of patients who switched to subsequent antineoplastic treatment was higher in the OC arm compared to the RCT arm (56.3% vs 52.9%; Table 3) during the whole follow-up period. Among patients who switched, the median time to treatment switch was shorter in the OC arm compared to pooled RCT arm (5.45 vs 6.24 months; 55.8% vs 46.1% switched in the first 6 months). Differences in pre-specified confounders for treatment switching including age, histology, treatment type, and progression were observed. Specifically, we saw a higher percentage of switching among patients with squamous and progression events during the follow-up period (Table 4).

## Main analyses

A logistic model was fitted (Appendix Table 4) to account for imbalances between the RCT and OC arms on baseline characteristics and estimate the propensity score (PS). Then IPTW weights were calculated using the propensity scores estimated from the logistic model and we excluded a small percentage of patients (0.4%) with extreme weights (weight > 10) in the OC arm to avoid undesirable variability in estimates due to extremely large weights.[17] SMDs for patient variables were all below 0.1[18] after IPTW (Figure 1) suggesting balance achieved on the selected baseline characteristics through IPTW weighting.

Patients were artificially censored at the time of treatment switching, then the censoring mechanism was modeled via a Cox regression model and the probability of not being censored conditional on patient/clinical characteristics that were pre-specified was estimated (Table 4). The stabilized IPCW weights were calculated as the ratio of inverse of the probability of not being censored conditional on race only and the probability of not being censored conditional on the age, race, histology and progression. Here, different from the traditional stabilized weights, race was added to both the numerator and denominator to further increase the stability of the IPCW weight.[19] In order to make stable estimation and reduce variability, extreme weights were trimmed at 99th percentile for the OC arm and 98th percentile for the RCT arm. After implementing both IPTW and IPCW, SMDs on the majority of baseline and time-varying confounders were reduced to values below 0.1 (Figure 2), and slightly larger than 0.1 for race, tumor type, and histology.

After accounting for treatment setting assignment at baseline and treatment switching using IPTW-IPCW method, the HR estimated from the weighted Cox model was equal to 0.94 (95% CI: [0.77, 1.13]), which suggests comparable overall survival between the RCT and OC arms. Weighted Kaplan-Meier estimates of survival functions (Figure 3) overall were comparable, however there was crossing hazard between two arms. The two curves align well at months 7-14, while RCT performed better at month 0-6 and worse at month 15-23. The difference in median survival time between the two arms was small (9.9 month with 95% CI: [8.6, 12.3] for OC vs 10.9 month with 95% CI: [9.6, 12.5] for RCT). These results suggested that after accounting for imbalances of baseline characteristics and removing the confounding effects of treatment switching, patients in the OC arm had similar OS as those in the RCT arm.

## Sensitivity analyses

A sensitivity analysis was performed to analyze the entire follow-up time period (i.e., no truncation) for the RCT and OC arms. The HR was 0.93 (95% CI: [0.77, 1.13]), which was similar to the primary analysis results. However, there were wider confidence intervals for K-M curves after month 21 for both the RCT and OC arms (Figure 4) due to the low number of events.

## Supplemental analyses

In supplemental analysis, we performed an IPTW-only analysis that adjusted for baseline characteristics only by IPTW weighting but without IPCW. This is a commonly-used method in analyses of external control arms, resulting in a different estimand compared to the primary analysis. Although the HR was similar with primary analysis 0.92 (95% CI: [0.81, 1.05]), there was a larger discrepancy in Kaplan-Meier estimates between RCT and OC, especially during Month 6 and 14 compared to the primary analysis.

## Discussion

In this study we applied the EF and TTF to define a precise scientific question in comparative-effectiveness research. As a case study to assess the feasibility of using observational data to construct an external control arm to single arm trials, we conducted a retrospective cohort study to compare OS among patients with metastatic NSCLC exposed to front-line chemotherapy in RCTs versus routine clinical practice settings, while accounting for differences in subsequent treatments between these settings. To achieve this objective, we pooled clinical trial patients from the control arms of three RCTs (IMPOWER 130, 131, and 132) and developed an OC cohort derived from de-identified EHR data obtained from routine clinical practice. OS was compared between the two arms, assuming a hypothetical scenario wherein patients in neither setting received subsequent therapy after the first-line chemotherapy. We found no difference in OS between the two arms. When accounting for baseline confounding as well as differences in patterns of subsequent treatments in clinical trial and routine clinical practice care patients, the long-term outcome of first-line treatment for patients with metastatic NSCLC is similar.

Our approach attempts to clarify the causal contrast of interest by combining elements of the EF and TTF. The EF and TTF serve complementary purposes in answering the scientific question. As formulated by Hernán and Robins, the TTF ensures that an appropriate comparative study is designed to help estimate the causal effect from the observed data[3]. While the causal contrast can be specified within the TTF, the EF adds clarity to the causal contrast through the explicit consideration of intercurrent events (i.e., events occurring post-baseline that can affect the assessment of treatment effects). Combined, the EF and TTF improve the transparency in the: 1) target of estimation (causal contrast), 2) assumptions and data needed to identify the causal contrast, and 3) limitations of available data.

To our knowledge, there are a limited number of studies that combine the EF and TTF. Recently, Hampson et al (2022)[20] combined the EF and TTF using routine clinical care data to generate an external control arm. The approach described in our study adds to the limited number of use cases by accounting for a scenario where patterns of subsequent treatments are different between the sources of clinical trial and routine clinical care. We anticipate that many researchers will likely encounter this scenario in applications involving real-world external controls. Our study, unlike other studies, also illustrates the iterative nature of specifying an estimand. In practice, such iteration allows a comprehensive and transparent dialogue among stakeholders to reach consensus on the scientific question and its tractability given the available data (i.e., discern the identifiability of the estimand).

Strengths of this study include the combination of the EF and TTF, as well as large sample size with extensive follow-up and a high proportion of patients with an event of interest. Furthermore, death events were ascertained with high accuracy in both RCT and observational data settings.[21] Lastly, model diagnostics indicated that the weights from IPTW and IPCW induced a good balance in the measured baseline and post-baseline confounders.

Limitations of the study include that because data were pooled from disparate sources, full information was not available for all possible confounders. For example, there was limited capture of comorbidities, sites of metastasis, and smoking status within the OC arm compared to the RCT arm. The assumption of no unmeasured confounders underlies both IPCW (i.e., baseline as well as time-varying covariates jointly predicting treatment switch and outcome;[22] as well as IPTW (i.e., baseline covariates jointly predicting treatment setting and outcome). About 43% (Table 2) of the patients in routine clinical care included in our study had missing ECOG-PS at the start of front-line therapy, some of whom may have had an ECOG-PS value above 1. For context, among adults with NSCLC who received first-line chemotherapy in the real-world setting, 13.6% had an ECOG-PS greater than 1 (Appendix Table 2). A second limitation was that the definition of time-zero differed across the RCT and OC arms. Time-zero was the date of randomization in the clinical trials as compared to the date of treatment initiation in the routine clinical practice care cohort. The impact is believed to be small given that typically treatment was initiated within a few days post-randomization. A third limitation is that patients in IMpower trials were global while patients in the OC arm were from the United State only. Although we account for potential patient confounders in our models, there could be residual confounding effects due to the difference in the region. A final limitation was that we pooled data from the control arms of the RCTs and hence assumed negligible heterogeneity in outcomes among the three clinical-trial cohorts. However, we believe trial heterogeneity posed little bias risk to our study because the survival estimates for each of the three trials were similar (Appendix Figure 1).

In conclusion, this study demonstrated that combining the EF and TTF approaches can improve the rigor in the design and analysis of comparative effectiveness studies, including retrospective observational studies. The EF approach alone does not suffice in specifying a study design, and the TTF alone can leave ambiguity in the inferential target. The combination of the two frameworks should be considered more often by researchers.

## Acknowledgements

The authors would like to thank Cody Patton and Hannah Gilham of Flatiron Health, Inc. for publication management support, as well as Somnath Sarkar and Meghna Samant of Flatiron Health, Inc for their scientific input and support of this research.

# Figures

## Figure 1. Covariate balance after IPTW

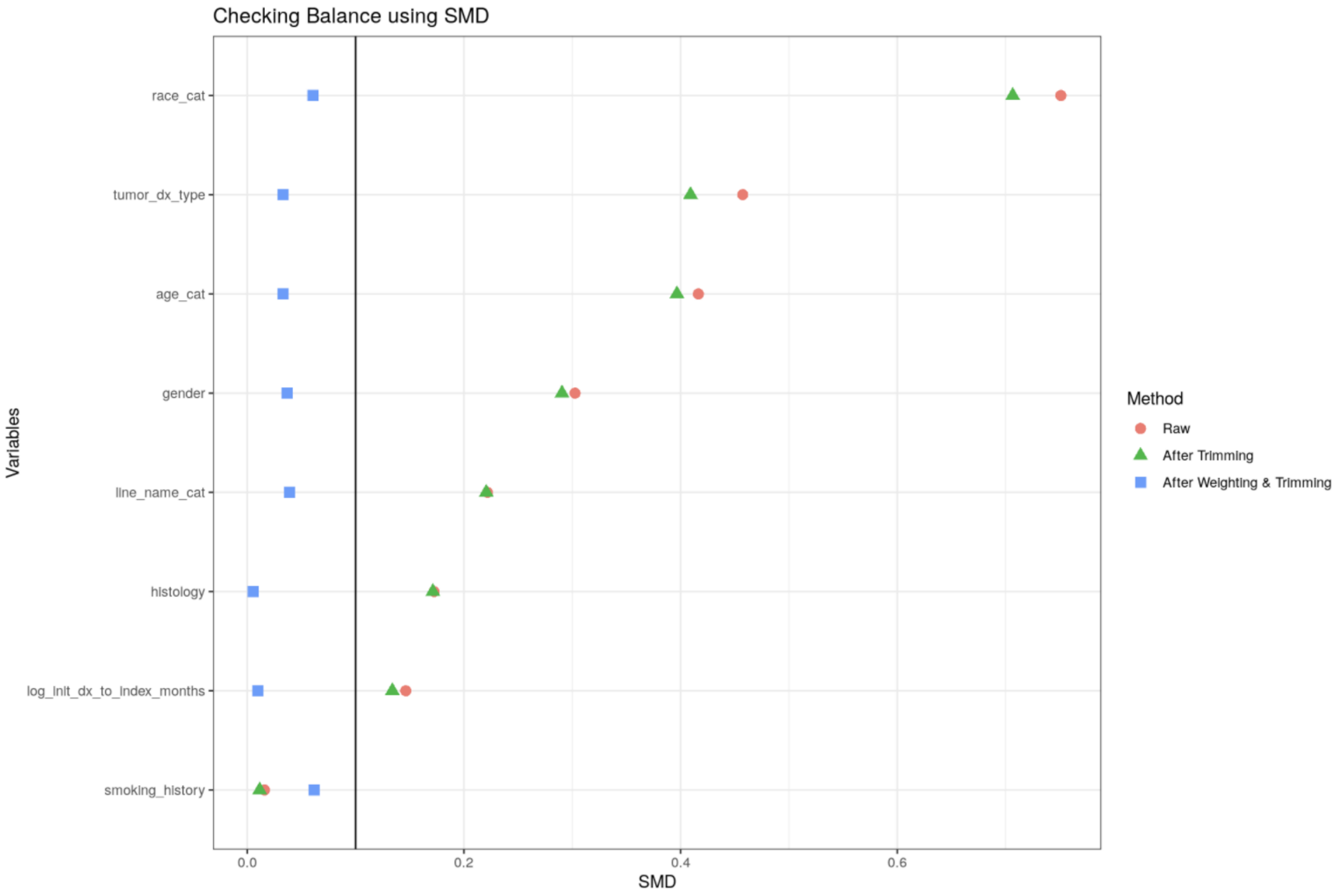

Figure 2. Covariate balance after IPTW and IPCW

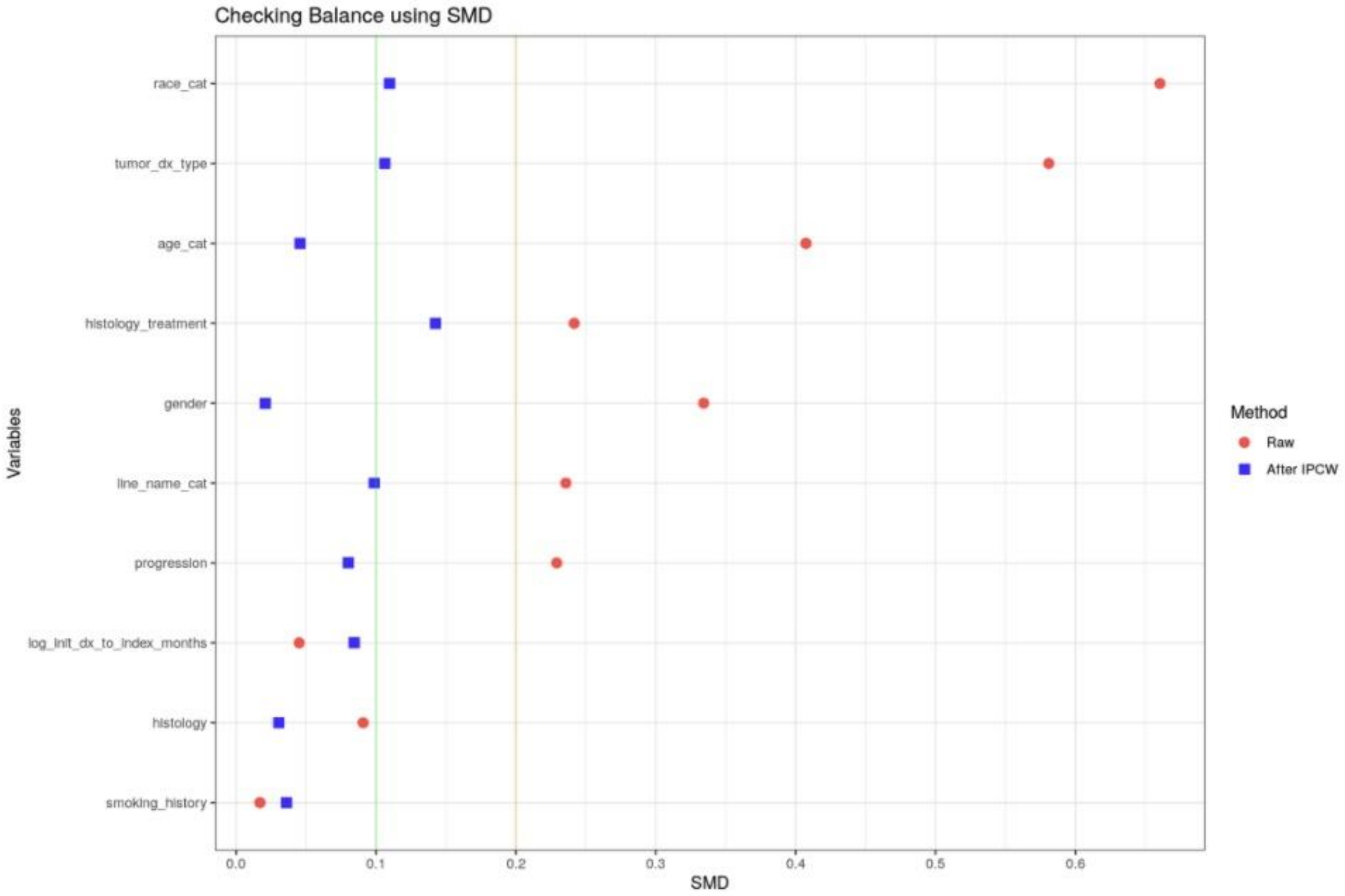

Figure 3. IPTW-IPCW weighted Kaplan-Meier curves

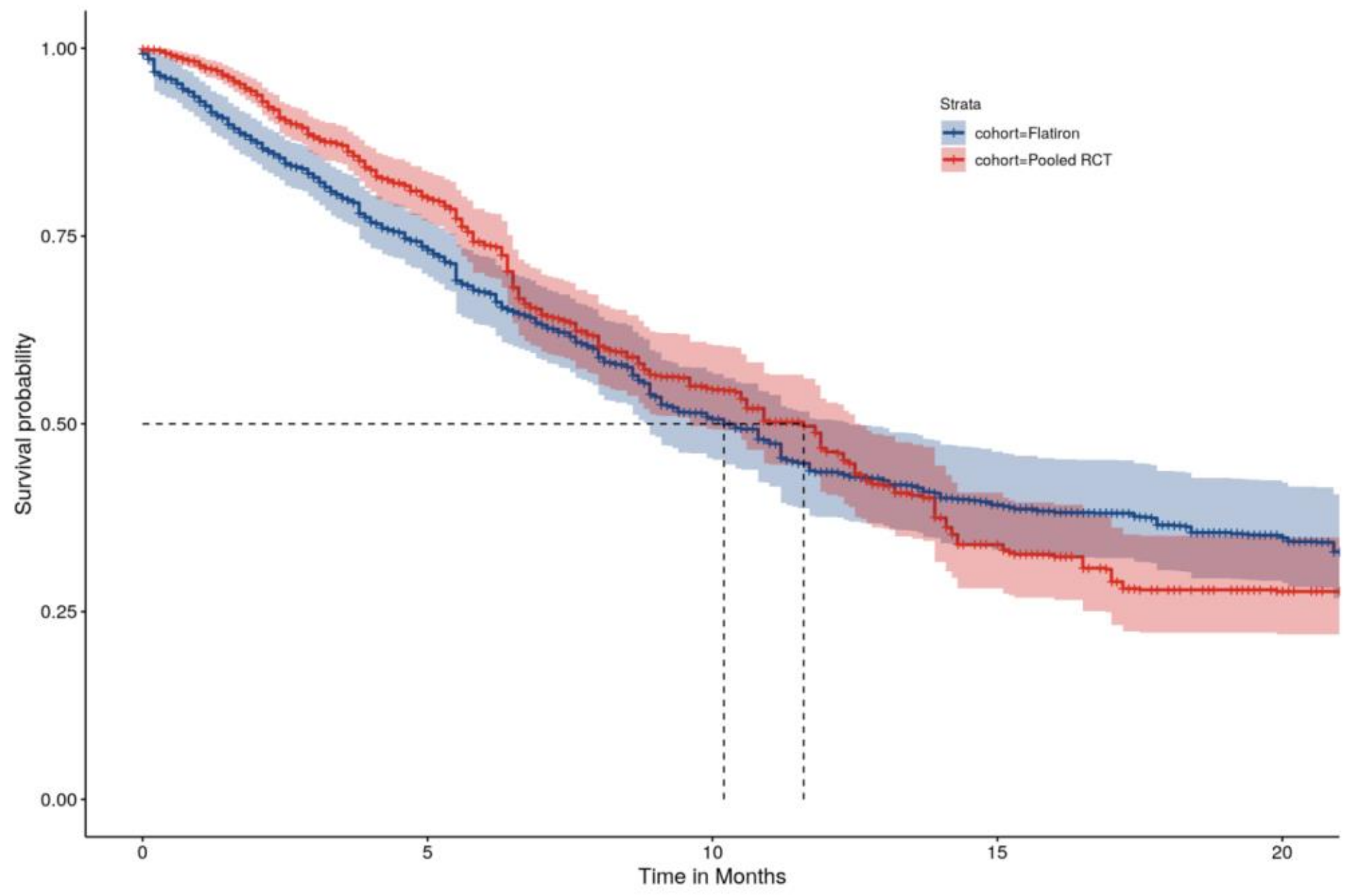

Figure 4. IPTW-IPCW weighted Kaplan-Meier survival function estimates without truncating the follow-up time

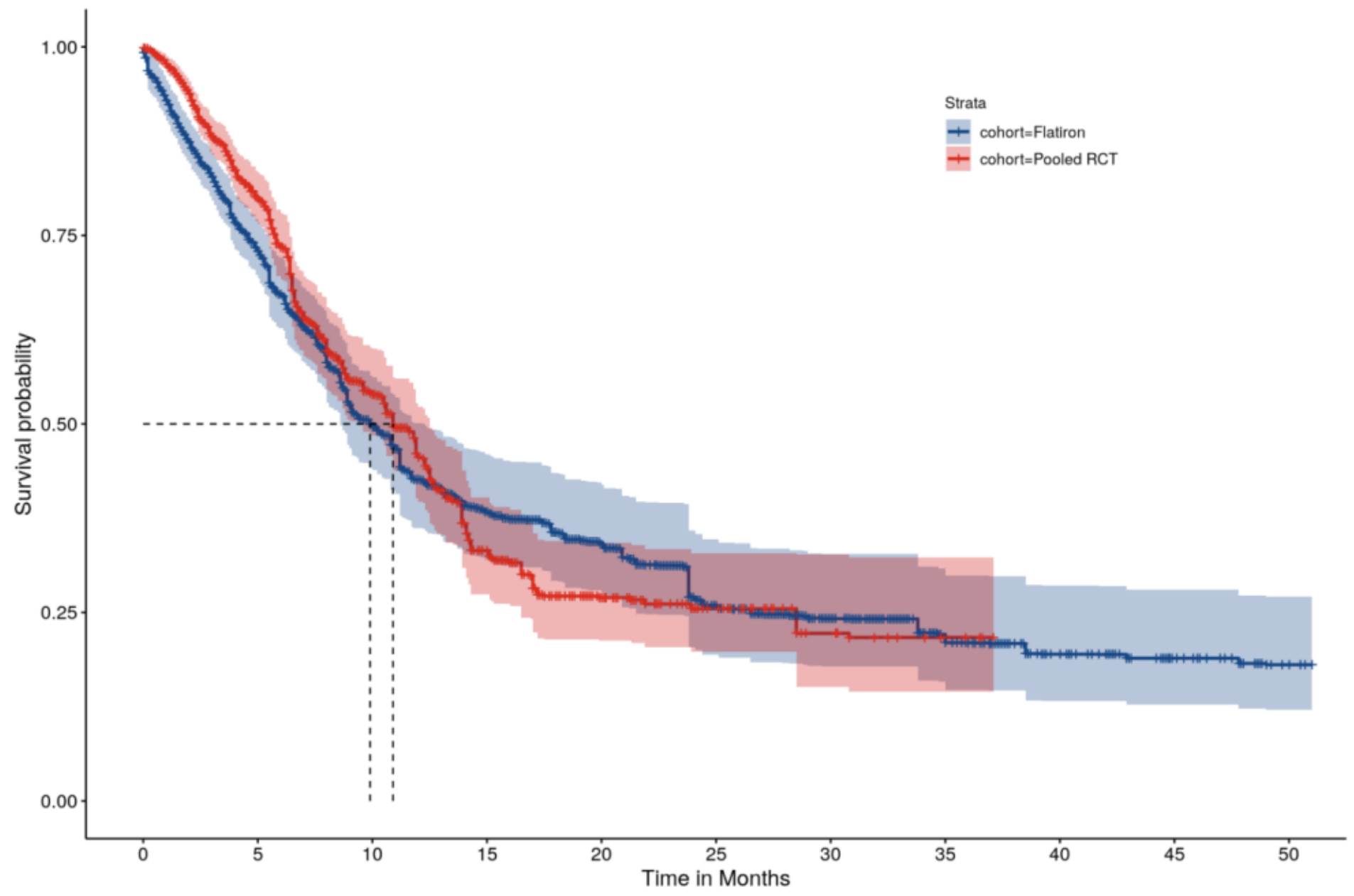

Figure 5. IPTW weighted Kaplan-Meier curves

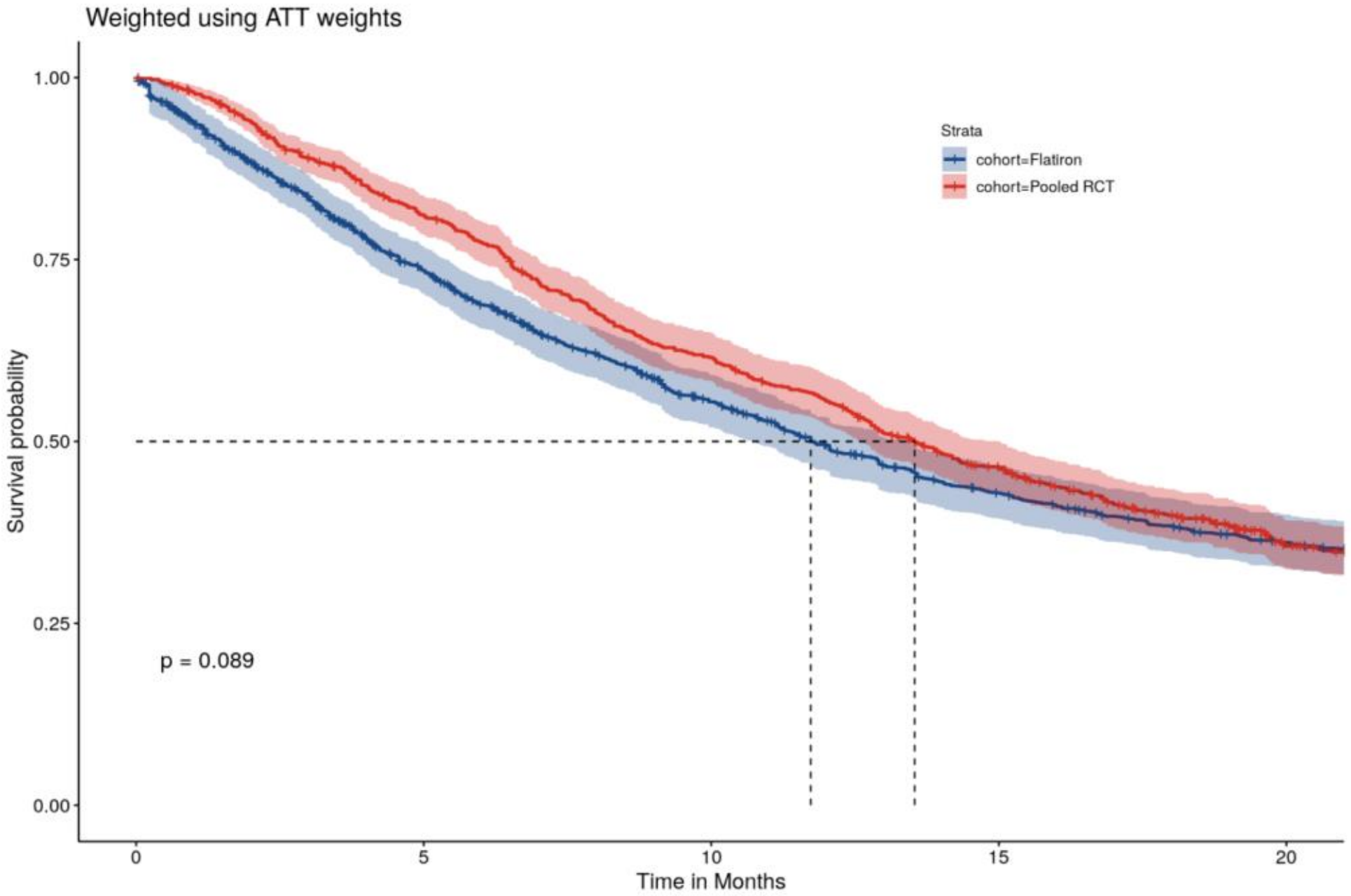

# Tables

Table 1. EF/TTF attributes based on the scientific research question

| ***Scientific research question*** | | | |
|---|---|---|---|
| Would there be a difference in overall survival (OS) between patients with metastatic NSCLC receiving front-line chemotherapy (control arms) in IMpower trials (130, 131 and 132) vs patients with metastatic NSCLC who received front-line chemotherapy as part of routine care, had patients not received a subsequent therapy? | | | |
| ***EF/TTF Attributes*** | **Target trial** | **Emulation of the target trial** | **Assumptions** |
| **Target population/Eligibility criteria** | Metastatic squamous and non-squamous NSCLC patients, 18 years of age or older, with ECOG PS 0,1 and with adequate hematological and end-organ function. | Same as the target trial for the RCT arm, with some assumptions for the OC arm. | Observational data does not perfectly emulate the trial I/E criteria. We attempt to define the study cohort that best approximates the target population by including additional rules.<br>• Time window for the eligibility assessment (ECOG PS, lab values, biomarker) |
| | The population is defined through the common inclusion and exclusion (I/E) criteria of IMpower130, 131 and 132 (limited to those criteria applicable retrospectively to observational data). To align the I/E criteria of the 3 trials, and to reflect the targeted population treated with 1L chemotherapy, patients with a sensitizing mutation in the EGFR gene or an ALK fusion oncogene were excluded | | • How to handle missing values (ECOG PS, lab values, biomarker)<br>○ Excluding patients with missing value may introduce selection bias<br>• Rules to account for difference between trial structured visits and routine clinical care<br>○ E.g., Patients with structured activity within 90 days of advanced diagnosis |
| **Treatment/Treatment strategies** | The investigational group (pooled trial control arms) and the observational comparator (OC) groups received the following chemotherapies:<br><br>Patients with non-squamous NSCLC:<br>-Pemetrexed + cisplatin / carboplatin;<br>-nab-paclitaxel/paclitaxel + carboplatin*<br><br>Patients with squamous NSCLC:<br>-nab-paclitaxel/paclitaxel + carboplatin* | Same as the target trial with some assumptions for both arms | Assumption on treatment:<br>● For this study we assume equivalence of nab-paclitaxel and paclitaxel. However, the two molecules are known to have different safety profiles. The decision to include paclitaxel was to limit treatment assignment bias since nab-paclitaxel is not the standard of care in the real world while it was adopted in IMpower trials. |
| | The investigational group received care according to the trial protocol, whereas the comparator group received care according to routine clinical practice. | | |
| **Endpoint/Outcomes** | Overall survival (OS) | Same as the target trial. | None.<br>The validity of the real-world OS (rwOS) from Flatiron Health has been demonstrated (Zhang et al 2021) against clinical trial OS as the gold standard to capture death occurrence. For this reason, in this study we refer to OS and not to rwOS for routine clinical practice. |
| **Intercurrent events (IE) and strategy/Causal contrast** | **IE:** Receipt of any subsequent cancer therapy<br>**Strategy:** hypothetical | Same as the target trial. | None. |
| | **Causal contrast**: Per-protocol effect of adhering to treatment after initiation. Receipt of any subsequent cancer therapy<br>is a deviation from the study protocol. | | |
| **Population-level Summary/analysis plan** | Hazard ratio (HR) with 95% confidence interval (CI) | Same as the target trial. | None. |
| **Assignment procedures** | Participants were randomly assigned to one of the two treatment settings | Randomization is emulated by weighting observations for the inverse probability of treatment setting | <u>Clinical assumptions</u><br>Treatment setting assignment was assumed to be conditional on the following baseline covariates:<br>• Age, gender, race, metastatic tumor type (de novo Stage IV/recurrent disease), time from initial diagnosis to index date, smoking history, histology and treatment type. |
| | | assignment following some assumptions | <u>Statistical assumptions</u><br>Statistical assumptions include consistency, conditional exchangeability, positivity and correct model specification. These are explained in the text. |
| **Start/end follow-up** | Start of follow-up occurs at the time when the treatment is assigned (i.e., when eligibility are met)<br><br>End of follow-up is reported in Appendix Table 1 | Same as target trial.<br><br>To emulate the start of follow up for the OC arm, some assumptions are needed. | For the OC arm, the actual start of follow-up occurs at the time when the treatment is initiated (dose 1 cycle 1).<br>The risk of comparing different time zero is to introduce immortal time bias. This cannot be quantified. The primary estimate is unbiased if the following assumptions are met.<br><br>Assumptions in the OC:<br>● There are no reasons for a patient to not initiate treatment other than death once assigned to treatment. |
| | | | <ul><li>Death is unlikely to have occurred in between assignment and start of treatment because we assume:<ul><li>The time between assignment and start of therapy is short</li><li>mNSCLC is a disease with no rapid course in first line</li></ul></li></ul>No assumption for RCT. We verified that:<ul><li>All patients assigned to treatment started treatment</li><li>Median time between assignment and start of therapy was 2 days</li></ul> |

I/E: inclusion and exclusion

OC: observational comparator

OS: overall survival

rwOS: real-world overall survival

HR: hazard ratio

CI: confidence interval

Table 2. Baseline characteristics

| Variable | Categories | RCT arm<br>N=849 | OC arm<br>N=3340 | P-value |
|---|---|---|---|---|
| Age group (years), n (%) | < 65 | 435 (51.2) | 1222 (36.6) | <0.001 |
| | ≥ 65 and < 75 | 322 (37.9) | 1268 (38.0) | |
| | ≥ 75 | 92 (10.8) | 850 (25.4) | |
| Gender, n (%) | Female | 248 (29.2) | 1457 (43.6) | <0.001 |
| Race, n (%) | Asian | 105 (12.4) | 46 (1.4) | <0.001 |
| | White | 699 (82.3) | 2373 (71.0) | |
| | Other | 45 (5.3) | 921 (27.6) | |
| ECOG-PS, n (%) | 0 | 314 (37.0) | 714 (21.4) | <0.001 |
| | 1 | 532 (62.7) | 1179 (35.3) | |
| | Unknown | 2 (0.2) | 1447 (43.3) | |
| Tumor diagnosis type, n (%) | De novo Stage IV | 706 (83.2) | 2118 (63.4) | <0.001 |
| | Recurrent disease | 143 (16.8) | 1221 (36.6) | |
| Smoking history, n (%) | No | 69 (8.1) | 257 (7.7) | 0.177 |
| | Yes | 780 (91.9) | 3070 (91.9) | |
| | Unknown | 0 (0.0) | 13 (0.4) | |
| Histology, n (%) | Non-squamous | 509 (60.0) | 2278 (68.2) | <0.001 |
| | Squamous | 340 (40.0) | 1062 (31.8) | |
| Time from initial diagnosis to index date (months), (median [IQR]) | | 1.41 [0.92, 2.89] | 1.25 [0.79, 2.27] | <0.001 |
| Treatment, n (%) | Carboplatin+Pacli/Nab-pacli | 568 (66.9) | 1877 (56.2) | <0.001 |
| | Platinum+Pemetrexed | 281 (33.1) | 1463 (43.8) | |

[a] The “unknown” category was not considered for SMD calculation.

[b] ECOG-PS variable was not included in the propensity score model because of the high proportion of missing values

Table 3. Characteristics of intercurrent events

| | RCT | OC |
|---|---|---|
| Number of patients | 849 | 3340 |
| Median (95% CI) follow-up time, months | 26.5 (19.9-28.8) | 35.6 (29.4-43) |
| Switch to subsequent therapy (any), n(%) | 449 (52.9%) | 1881 (56.3%) |
| Median (IQR) time to switch (among patients who switched), months | 6.24 (4.27-9.69) | 5.45 (3.12-9.43) |
| Number of patients who switched prior to 6 months / Number of patients who ever switched, n (%) | 207/449 (46.1) | 1049/1881 (55.8) |

Table 4. Baseline and clinical characteristics among patients who switched treatment and who did not switch treatment

| Variable | Category, n (%) | RCT | | OC | |
|---|---|---|---|---|---|
| | | Patients who switched treatment N=449 (52.9%) | Patients who did not switch treatment N=400 (47.1%) | Patients who switched treatment N=1881 (56.3%) | Patients who did not switch treatment N=1459 (43.7%) |
| Age | < 65 | 227 (50.6) | 207 (51.9) | 708 (37.6) | 514 (35.2) |
| | 65 - 75 | 179 (39.9) | 143 (35.8) | 717 (38.1) | 551 (37.8) |
| | ≥ 75 | 43 (9.6) | 49 (12.3) | 456 (24.2) | 394 (27.0) |
| Histology | Non-squamous | 251 (55.9) | 258 (64.5) | 1287 (68.4) | 991 (67.9) |
| | Squamous | 198 (44.1) | 142 (35.5) | 594 (31.6) | 468 (32.1) |
| Treatment | Carboplatin+Pacli/Nab-pacli | 287 (63.9) | 281 (58.5) | 1034 (55.0) | 843 (57.8) |
| | Platinum+Pemetrexed | 162 (36.1) | 119 (41.5) | 847 (45.0) | 616 (42.2) |
| Progression during the follow-up[a] | Yes | 390 (86.9) | 230 (57.5) | 1360 (72.3) | 397 (27.2) |
| | No | 59 (13.1) | 170 (42.5) | 521 (27.7) | 1062 (72.8) |

[a] follow-up is up to switch or, in absence of switch until last activity before study end date (end date for the specific data source)

# Appendix Figures

Figure 1: Kaplan-Meier plot showing the comparison of OS with the trials were considered separately

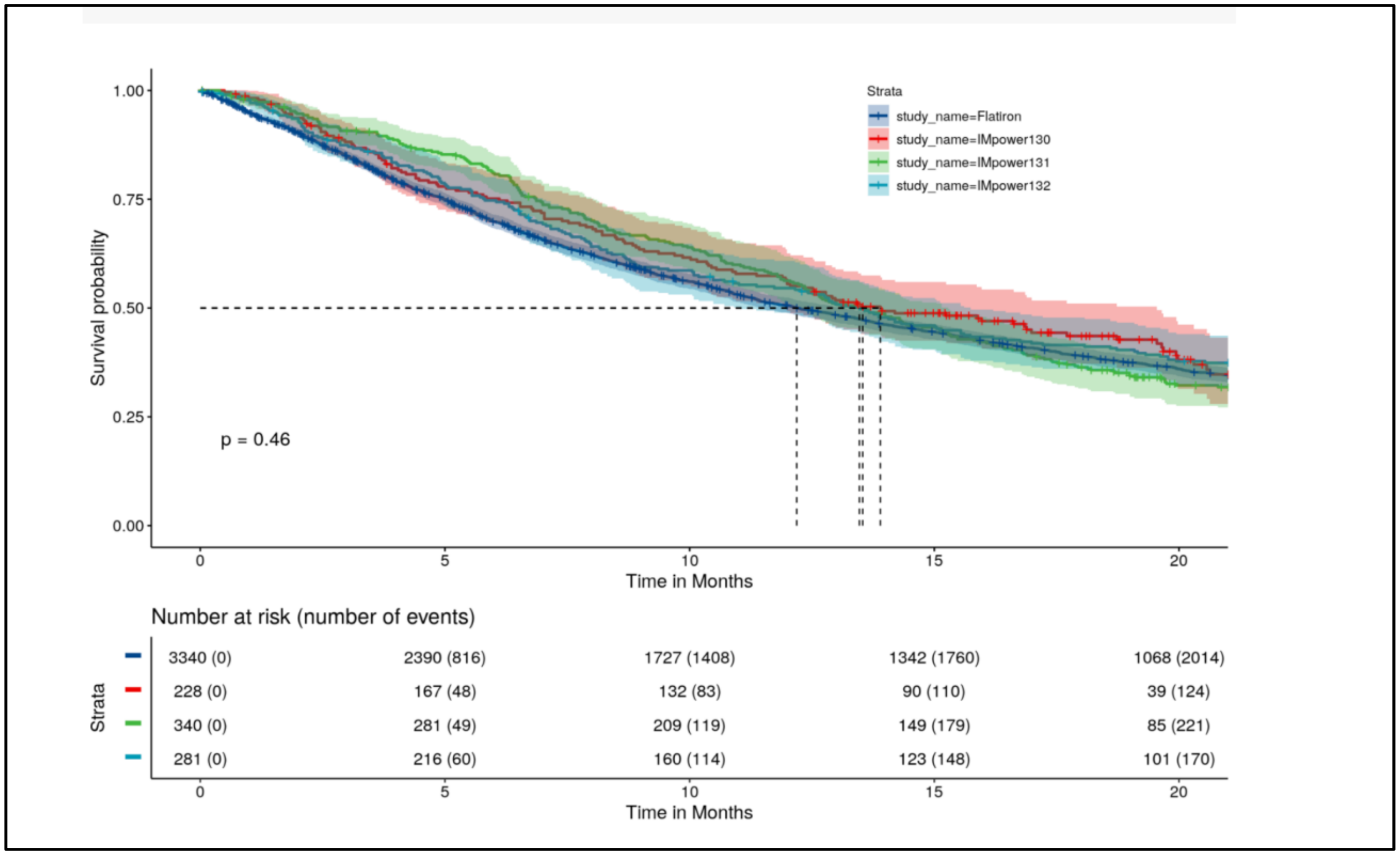

# Appendix Tables

Table 1. Alignment of study time window between RCT and OC arms

| Study | Enrollment start date | Enrollment end date | Primary completion date |
|---|---|---|---|
| IMpower130 | Apr 16, 2015 | Feb 13, 2017 | Mar 15, 2018 |
| IMpower131 | Jun 11, 2015 | Mar 28, 2017 | Oct 3, 2018 |
| IMpower132 | Apr 7, 2016 | May 31, 2017 | July 18, 2019 |
| **Pooled RCT arm** | **Apr 16, 2015** | **May 31, 2017** | **July 18, 2019** |
| **OC arm [a]** | **Apr 16, 2015 [b]** | **May 31, 2017** | **July 18, 2019** |

[a] OC arm covers the entire enrollment and follow-up period of the three trials

[b] Treatment initiation date

Table 2. OC attrition table

| Criteria | Number of patients |
|---|---|
| Total patients in the cohort (Dec 2020) | 66520 |
| Line start date is between ‘2015-04-16’ and ‘2017-05-31’ | 15009 |
| 1L after advanced diagnosis | 12310 |
| Patients with multiple primary tumors excluded | 12190 |
| Age >= 18 years-old | 12190 |
| ECOG PS 0, 1, or missing at baseline [a] | 10532 |
| Histology known | 10052 |
| With structured activities within 90 days of advanced diagnosis | 9255 |
| Without abnormal labs or missing within a -30 to +7 window around 1L start date | 8412 |
| Without EGFR mutation or missing at any time up to 7 days after 1L start | 7493 |
| Without ALK mut or missing at any time up to 7 days after 1L start | 7313 |
| No exposure to chemotherapy within prior six months | 7004 |
| No exposure to prior CIT | 6984 |
| With the regimens of interest in 1L | 3340 |

[a] The closest ECOG value within a -30 to +7 window around 1L start date was used, taking the highest if multiple observations occurred on the same day

Table 3. Baseline characteristics by trial

| Variable | Categories | RCT | | | | OC N=3340 | P-value | SMD |
|---|---|---|---|---|---|---|---|---|
| | | Total N=849 | IMpower 130 N=228 | IMpower 131 N=340 | IMpower 132 N=281 | | | |
| Age group (years), n(%) | < 65 | 435 (51.2) | 114 (50.0) | 156 (45.9) | 164 (58.4) | 1222 (36.6) | <0.001 | 0.42 |
| | ≥ 65 and < 75 | 322 (37.9) | 83 (36.4) | 145 (42.6) | 94 (33.5) | 1268 (38.0) | | |
| | ≥ 75 | 92 (10.8) | 30 (13.2) | 39 (11.5) | 23 (8.2) | 850 (25.4) | | |
| Gender, n(%) | Female | 248 (29.2) | 94 (41.2) | 63 (18.5) | 91 (32.4) | 1457 (43.6) | <0.001 | 0.30 |
| Race, n(%) | Asian | 105 (12.4) | 3 (1.3) | 37 (10.9) | 65 (23.1) | 46 (1.4) | <0.001 | 0.75 |
| | White | 699 (82.3) | 210 (92.1) | 290 (85.3) | 199 (70.8) | 2373 (71.0) | | |
| | Other | 45 (5.3) | 15 (6.6) | 13 (3.8) | 17 (6.0) | 921 (27.6) | | |
| ECOG-PS, n(%) | 0 | 314 (37.0) | 91 (39.9) | 110 (32.4) | 113 (40.2) | 714 (21.4) | <0.001 | 0.05* |
| | 1 | 532 (62.7) | 136 (59.6) | 229 (67.4) | 167 (59.4) | 1179 (35.3) | | |
| | 2 | 1 (0.0) | 1 (0.4) | 0 (0.0) | 0 (0.0) | 0 (0.0) | | |
| | Unknown | 2 (0.2) | 0 (0.0) | 1 (0.3) | 1 (0.4) | 1447 (43.3) | | |
| Metastatic diagnosis, n(%) | De novo Stage IV | 706 (83.2) | 188 (82.5) | 270 (79.4) | 248 (88.3) | 2118 (63.4) | <0.001 | 0.46 |
| | Recurrent disease | 143 (16.8) | 40 (17.5) | 70 (20.6) | 33 (11.7) | 1221 (36.6) | | |
| | No | 69 (8.1) | 17 (7.5) | 24 (7.1) | 28 (10.0) | 257 (7.7) | 0.177 | 0.02 |
| Smoking history, n(%) | Yes | 780 (91.9) | 211 (92.5) | 316 (92.9) | 253 (90.0) | 3070 (91.9) | | |
| | Unknown | 0 (0.0) | 0 (0.0) | 0 (0.0) | 0 (0.0) | 13 (0.4) | | |
| Histology, n(%) | Non-squamous | 509 (60.0) | 228 (100.0) | 0 (0.0) | 281 (100.0) | 2278 (68.2) | <0.001 | 0.17 |
| | Squamous | 340 (40.0) | 0 (0.0) | 340 (100.0) | 0 (0.0) | 1062 (31.8) | | |
| Time from initial diagnosis to index date (months), (median [IQR]) | | 1.41 [0.92, 2.89] | 1.58 [1.08, 3.08] | 1.41 [0.91, 3.96] | 1.35 [0.92, 2.43] | 1.25 [0.79, 2.27] | <0.001 | 0.15 |
| Treatment, n(%) | Carboplatin +Pacli/Nab-pacli | 568 (66.9) | 228 (100.0) | 340 (100.0) | 0 (0.0) | 1877 (56.2) | <0.001 | 0.22 |
| | Platinum+Pemetrexed | 281 (33.1) | 0 (0.0) | 0 (0.0) | 281 (100.0) | 1463 (43.8) | | |

*ECOG-PS variable was not included in the propensity score model because of the high proportion of missing values.

Table 4. Parameter estimates from the logistic model

| Variable | Category | Parameter Estimate | Standard Error | P-value |
| --- | --- | --- | --- | --- |
| Intercept | | 0.999 | 0.264 | <0.001 |
| Age | 65-75 | -0.509 | 0.095 | <0.001 |
| | >75 | -1.342 | 0.135 | <0.001 |
| Gender | Male | 0.559 | 0.094 | <0.001 |
| Race | White | -2.094 | 0.201 | <0.001 |
| | Other | -3.959 | 0.251 | <0.001 |
| Smoking history | Yes | -0.183 | 0.166 | 0.269 |
| Tumor diagnosis type | Recurrent | -2.446 | 0.169 | <0.001 |
| Log-transformed time from initial diagnosis to index date (month) | | 0.840 | 0.065 | <0.001 |
| Histology | Squamous | 0.076 | 0.111 | 0.492 |
| Line name | Platinum+Pemetrexed | -0.832 | 0.113 | <0.001 |

Table 5. Detailed I/E criteria

| **IMpower130** | **IMpower131** | **IMpower132** | **OC** |
|---|---|---|---|
| ***Inclusion Criteria*** | | | |
| *Signed Informed Consent Form* | *Signed Informed Consent Form* | *Signed Informed Consent Form* | *Not implemented (Not relevant)* |
| *Male or female, 18 years of age or older* | *Male or female, 18 years of age or older* | *Male or female, 18 years of age or older* | *Implemented (Age at start of treatment)* |
| *ECOG performance status of 0 or 1* | *ECOG performance status of 0 or 1* | *ECOG performance status of 0 or 1* | *Implemented ECOG PS 0 or 1 or missing (The closest ECOG value within a -30 to +7 window around 1L start date will be used, taking the highest if multiple observations occur on same day)* |
| *Histologically or cytologically confirmed, Stage IV **non-squamous** NSCLC* | *Histologically or cytologically confirmed, Stage IV **squamous** NSCLC* | *Histologically or cytologically confirmed, Stage IV **non-squamous** NSCLC*<br>- *Patients with tumors of mixed non-small cell histology (i.e., squamous and non-squamous) were eligible if the major histological component appeared to be non-squamous.* | *Implemented (either squamous or non-squamous)* |
| *No prior treatment for Stage IV **non-squamous** NSCLC* | *No prior treatment for Stage IV **squamous** NSCLC* | *No prior treatment for Stage IV **non-squamous** NSCLC* | *Implemented (Patients with structured activity within 90 days of advanced diagnosis)* |
| *Patients who have received prior neo-adjuvant, adjuvant chemotherapy, or chemoradiotherapy with curative intent for non-metastatic disease must have experienced a treatment-free interval of at least 6 months from randomization since the last chemotherapy or chemoradiotherapy cycle.* | *Patients who have received prior neo-adjuvant, adjuvant chemotherapy,* ***radiotherapy****, or chemoradiotherapy with curative intent for non-metastatic disease must have experienced a treatment-free interval of at least 6 months from randomization since the last chemotherapy, radiotherapy, or chemoradiotherapy.* | *Patients who had received prior neo-adjuvant, adjuvant chemotherapy,* ***radiotherapy****, or chemoradiotherapy with curative intent for non-metastatic disease must have experienced a treatment-free interval of at least 6 months from randomization since the last dose of chemotherapy and/or radiotherapy* | *Implemented (Patients with documented antineoplastic medications 6 months prior to 1L start date, per medication administration and enhanced orals tables)* |
| *Patients with a history of treated asymptomatic CNS metastases are eligible, provided they* | *Patients with a history of treated asymptomatic CNS metastases are eligible, provided they* | *Patients with a history of treated asymptomatic CNS metastases were eligible, provided they met all of the* | *Not implemented (Not in EDM data model)* |
| *meet all of the following criteria:*<br>- *Only supratentorial and cerebellar metastases allowed (i.e., no metastases to midbrain, pons, medulla or spinal cord)*<br>- *No ongoing requirement for corticosteroids as therapy for CNS disease*<br>- *No stereotactic radiation within 7 days or whole-brain radiation within 14 days prior to* | *meet all of the following criteria:*<br>- *Only supratentorial and cerebellar metastases allowed (i.e., no metastases to midbrain, pons, medulla, or spinal cord)*<br>- *No ongoing requirement for corticosteroids as therapy for CNS disease*<br>- *No stereotactic radiation within 7 days or whole-brain radiation within 14 days prior to* | *following criteria:*<br>- *Only supratentorial and cerebellar metastases were allowed (i.e., no metastases to midbrain, pons, medulla, or spinal cord)*<br>- *No ongoing requirement for corticosteroids as therapy for CNS disease*<br>- *No stereotactic radiation within 7 days or whole-brain radiation within 14 days prior to* | |
| *randomization*<br>- *No evidence of interim progression between the completion of CNS-directed therapy and the screening radiographic study* | *randomization*<br>- *No evidence of interim progression between the completion of CNS-directed therapy and the screening radiographic study.* | *randomization*<br>- *No evidence of interim progression between the completion of CNS-directed therapy and the screening radiographic study* | |
| *NA* | *NA* | *Patients were to submit a pre-treatment tumor tissue sample (if available). If tumor tissue was not available (e.g., depleted for prior diagnostic testing), patients were still eligible.* | *Not implemented* |
| ***Known PD-L1 tumor*** | ***Known PD-L1 tumor*** | ***Known tumor PD-L1*** | *Not implemented* |
| ***status*** *as determined by an IHC assay performed by a central laboratory on previously obtained archival tumor tissue or tissue obtained from a biopsy at screening* | ***status*** *as determined by an IHC assay performed by a central laboratory on previously obtained archival tumor tissue or tissue obtained from a biopsy at screening.* | *expression status as determined by an IHC assay from other clinical studies* | *(Include all patients regardless of PD-L1 status)* |
| *Measurable disease, as defined by RECIST v1.1* | *Measurable disease, as defined by RECIST v1.1* | *Measurable disease, as defined by RECIST v1.1* | *Not implemented (Not in EDM data model)* |
| *Adequate hematologic and end-organ function, defined by the following laboratory results obtained within 14 days prior to randomization:* | *Adequate hematologic and end-organ function, defined by the following laboratory results obtained within 14 days prior to randomization:* | *Adequate hematologic and end-organ function, defined by the following laboratory results obtained within 14 days prior to randomization:* | *See the following 8 rows. Adequate hematological and end-organ function at start of index therapy. For all of the above labs, the closest result in the -28 to 0 window around 1L start will be used. Patients with no* |
| | | | *recorded results for a specific lab in that time window will not be excluded.* |
| - *ANC ≥1500 cells/µL without granulocyte colony-stimulating factor support* | - *ANC ≥1500 cells/µL without granulocyte colony-stimulating factor support* | - *ANC ≥1500 cells/µL without granulocyte colony-stimulating factor support* | *Implemented* |
| - *Lymphocyte count ≥500 cells/µL* | - *Lymphocyte count ≥500 cells/µL* | - *Lymphocyte count ≥500 cells/µL* | *Implemented* |
| - *Platelet count ≥100,000 cells/µL without transfusion* | - *Platelet count ≥100,000 cells/µL without transfusion* | - *Platelet count ≥100,000 cells/µL without transfusion* | *Implemented* |
| - *Hemoglobin ≥9.0 g/dL* | - *Hemoglobin ≥9.0 g/dL* | - *Hemoglobin ≥9.0 g/dL* | *Implemented* |
| - *INR or aPTT ≤1.5 x upper limit of normal (ULN)* | - *INR or aPTT ≤1.5 x upper limit of normal (ULN)* | - *INR or aPTT ≤1.5 x upper limit of normal (ULN)* | *Not Implemented (Highly missing)* |
| - *AST, ALT, and alkaline phosphatase ≤2.5 x ULN with the following exceptions: Patients with* | - *AST, ALT, and alkaline phosphatase ≤2.5 x ULN with the following exceptions: Patients with* | - *AST, ALT, and alkaline phosphatase ≤2.5 x ULN with the following exceptions: Patients with* | *Implemented (Same, though will not consider the exceptions)* |
| documented liver metastases: AST and/or ALT ≤5 x ULN; Patients with documented liver or bone metastases: alkaline phosphatase ≤5 x ULN | documented liver metastases: AST and/or ALT ≤5 x ULN; Patients with documented liver or bone metastases: alkaline phosphatase ≤5 x ULN | documented liver metastases: AST and/or ALT ≤5 x ULN; Patients with documented liver or bone metastases: alkaline phosphatase ≤5 x ULN | |
| - Serum bilirubin ≤1.5 x ULN | - Serum bilirubin ≤1.5 x ULN | - Serum bilirubin ≤**1.25** x ULN | Implemented (≤1.5 x ULN) |
| *N/A* | - *Calculated creatinine clearance (CrCl) ≥45 mL/min, or if using cisplatin, calculated CrCl ≥60 mL/min* | - *Calculated creatinine clearance (CrCl) ≥45 mL/min, or if using cisplatin, calculated CrCl ≥60 mL/min* | *Not implemented* |
| *NA* | *NA* | *For patients enrolled in the extended China enrollment phase: current resident of mainland China, Hong Kong, or Taiwan and of Chinese ancestry* | *Not implemented*<br>*(Not in EDM data model)* |
| *For female patients of* | *For female patients of* | *For women of* | *Not implemented* |
| *childbearing potential and male patients with partners of childbearing potential, agreement (by patient and/or partner) to use a highly effective form(s) of contraception that results in a low failure rate [< 1% per year] when used consistently and correctly, and to continue its use for* ***90 days*** *after the last dose of atezolizumab or for 30 days after the last dose of nab-paclitaxel, whichever is later. For male patients with* | *childbearing potential, agreement (by patient and/or partner) to use a highly effective form(s) of contraception that results in a low failure rate (< 1% per year) when used consistently and correctly, and to continue its use for* ***5 months*** *after the last dose of atezolizumab, for 30 days after the last dose of nab-paclitaxel, or for 6 months after the last dose of paclitaxel, whichever is later. Women must refrain from donating eggs* | *childbearing potential: agreement to remain abstinent (refrain from heterosexual intercourse) or use contraceptive methods that result in a failure rate of  1% per year during the treatment period and for at least 5 months after the last dose of atezolizumab or 6 months after the last dose of cisplatin. A woman was considered to be of childbearing potential if she was postmenarcheal, had not* | *(Not in EDM data model)* |
| *female partners of childbearing potential, agreement (by patient and/or partner) to use a highly effective form(s) of contraception that results in a low failure rate [< 1% per year] when used consistently and correctly, and to continue its use for 90 days after the last dose of atezolizumab or for 6 months after the last dose of nab-paclitaxel, whichever is later. Such methods include: combined(estrogen and progestogen containing)* | *during this same period. For male patients with female partners of childbearing potential, agreement (by patient and/or partner) to use a highly effective form(s) of contraception that results in a low failure rate (< 1% per year) when used consistently and correctly, and to continue its use for 6 months after the last dose of nab-paclitaxel, paclitaxel, and/or carboplatin. Such methods include combined (estrogen and* | *reached a postmenopausal state ( 12 continuous months of amenorrhea with no identified cause other than menopause), and had not undergone surgical sterilization (removal of ovaries and/or uterus). Examples of non-hormonal contraceptive methods with a failure rate of < 1% per year included bilateral tubal ligation, male sterilization, established, proper use of hormonal* | |
| *hormonal contraception, progestogen-only hormonal contraception associated with inhibition of ovulation together with another additional barrier method always containing a spermicide, intrauterine device (IUD), intrauterine hormone -releasing system (IUS), bilateral tubal occlusion or vasectomized partner (on the understanding that this is the only one partner during the whole study duration), and sexual abstinence. Male patients should not* | *progestogen containing) hormonal contraception, progestogen-only hormonal contraception associated with inhibition of ovulation together with another additional barrier method always containing a spermicide, intrauterine device (IUD), intrauterine hormone-releasing system (IUS), bilateral tubal occlusion or vasectomized partner (on the understanding that this is the only one partner during the whole study duration), and sexual abstinence. Men* | *contraceptives that inhibit ovulation, hormone-releasing intrauterine devices, and copper intrauterine devices. The reliability of sexual abstinence was to be evaluated in relation to the duration of the clinical trial and the preferred and usual lifestyle of the patient. Periodic abstinence (e.g., calendar, ovulation, symptothermal, or postovulation methods) and withdrawal were not* | |
| *donate sperm during this study and for at least 6 months after the last dose of nab-paclitaxel.* | *must refrain from donating sperm during the study and for 6 months after the last dose of nab-paclitaxel, paclitaxel, and/or carboplatin, whichever is latest.* | *acceptable methods of contraception.*<br><br>*For men: agreement to remain abstinent (refrain from heterosexual intercourse) or use contraceptive measures and agreement to refrain from donating sperm, as defined below:*<br>*With partners of childbearing potential, men were to remain abstinent or use a condom plus an additional contraceptive method that together resulted in a failure rate* | |
| *Oral contraception should always be combined with an additional contraceptive method because of a potential interaction with the study drug. The same rules are valid for male patients involved in this clinical study if they have a partner of* | *Oral contraception should always be combined with an additional contraceptive method because of a potential interaction with the study drug. The same rules are valid for male patients involved in this clinical trial if they have a partner of* | | *Not implemented*<br>*(Not in EDM data model)* |
| *childbirth potential. Male patients must always use a condom.* | *childbearing potential. Male patients must always use a condom.* | *of 1% per year during the chemotherapy treatment period and for at least 6 months after the last dose of chemotherapy. With pregnant partners, men were to remain abstinent or use a condom during the chemotherapy treatment period and for at least 6 months after the last dose of chemotherapy. The reliability of sexual abstinence was to be evaluated in relation to the duration of the clinical trial and the* | |
| *Women who are not postmenopausal (≥12 months of non-therapy-induced amenorrhea) or surgically sterile must have a negative serum pregnancy test result within 14 days prior to initiation of study drug.* | *Women who are not postmenopausal (≥12 months of non-therapy induced amenorrhea) or surgically sterile must have a negative serum pregnancy test result within 14 days prior to initiation of study drug.* | | *Not implemented (Not in EDM data model)* |
| | | *preferred and usual lifestyle of the patient. Periodic abstinence (e.g., calendar, ovulation, symptothermal, or postovulation methods) and withdrawal were not acceptable methods of contraception.* | |
| | | | |
| ***Exclusion*** | | | |
| ***Cancer-Specific Exclusions*** | | | |
| *Known tumor PD-L1 expression status as determined by an IHC assay from other clinical* | *Known tumor PD-L1 expression status as determined by an IHC assay from other clinical* | *Known tumor PD-L1 expression status as determined by an IHC assay from other clinical* | *Not implemented* |
| *studies* | *studies* | *studies* | |
| *N/A* | *N/A* | *Patients with a sensitizing mutation in the EGFR gene or an ALK fusion oncogene* | *Implemented* |
| *Active or untreated CNS metastases as determined by CT or MRI evaluation during screening and prior radiographic assessments* | *Active or untreated CNS metastases as determined by CT or MRI evaluation during screening and prior radiographic assessments* | *Active or untreated CNS metastases as determined by CT or MRI evaluation during screening and prior radiographic assessments* | *Not implemented (Not in EDM data model)* |
| *Spinal cord compression not definitively treated with surgery and/or radiation or previously diagnosed and treated spinal cord compression* | *Spinal cord compression not definitively treated with surgery and/or radiation or previously diagnosed and treated spinal cord compression* | *Spinal cord compression not definitively treated with surgery and/or radiation or previously diagnosed and treated spinal cord compression* | *Not implemented (Not in EDM data model)* |
| *without evidence that disease has been clinically stable for >2 weeks prior to randomization* | *without evidence that disease has been clinically stable for >2 weeks prior to randomization* | *without evidence that disease had been clinically stable for ≥ 2 weeks prior to randomization* | |
| *Leptomeningeal disease* | *Leptomeningeal disease* | *Leptomeningeal disease* | *Not implemented (Not in EDM data model)* |
| *Uncontrolled tumor-related pain* | *Uncontrolled tumor-related pain* | *Uncontrolled tumor-related pain* | *Not implemented (Not in EDM data model)* |
| *Uncontrolled pleural effusion, pericardial effusion, or ascites requiring recurrent drainage procedures (once monthly or more frequently)* | *Uncontrolled pleural effusion, pericardial effusion, or ascites requiring recurrent drainage procedures (once monthly or more frequently)* | *Uncontrolled pleural effusion, pericardial effusion, or ascites requiring recurrent drainage procedures (once monthly or more frequently)* | *Not implemented (Not in EDM data model)* |
| *Uncontrolled or* | *Uncontrolled or* | *Uncontrolled or* | *Implemented* |
| *symptomatic hypercalcemia (>1.5 mmol/L ionized calcium or Ca >12 mg/dL or corrected serum calcium >ULN)* | *symptomatic hypercalcemia (>1.5 mmol/L ionized calcium or Ca >12 mg/dL or corrected serum calcium >ULN)* | *symptomatic hypercalcemia (>1.5 mmol/L ionized calcium or Ca >12 mg/dL or corrected serum calcium >ULN)* | *(Same)* |
| *Malignancies other than NSCLC within 5 years prior to randomization, with the exception of those with a negligible risk of metastasis or death treated with expected curative outcome* | *Malignancies other than NSCLC within 5 years prior to randomization, with the exception of those with a negligible risk of metastasis or death treated with expected curative outcome* | *Malignancies other than NSCLC within 5 years prior to randomization, with the exception of those with a negligible risk of metastasis or death treated with expected curative outcome* | *Not implemented*<br>*(Not in EDM data model)* |
| | | | |
| ***General Medical Exclusions*** | | | |
| *Women who are pregnant, lactating, or intending to become pregnant during the study* | *Women who are pregnant, lactating or intending to become pregnant during the study* | *Women who are pregnant, lactating or intending to become pregnant during the study* | *Not implemented (Not in EDM data model)* |
| *History of severe allergic, anaphylactic, or other hypersensitivity reactions to chimeric or humanized antibodies or fusion proteins* | *History of severe allergic, anaphylactic, or other hypersensitivity reactions to chimeric or humanized antibodies or fusion proteins* | *History of severe allergic, anaphylactic, or other hypersensitivity reactions to chimeric or humanized antibodies or fusion proteins* | *Not implemented (Not in EDM data model)* |
| *Known hypersensitivity or allergy to biopharmaceuticals produced in Chinese hamster ovary cells or any component of the atezolizumab formulation* | *Known hypersensitivity or allergy to biopharmaceuticals produced in Chinese hamster ovary cells or any component of the atezolizumab formulation* | *Known hypersensitivity or allergy to biopharmaceuticals produced in Chinese hamster ovary cells or any component of the atezolizumab formulation* | *Not implemented (Not in EDM data model)* |
| *History of autoimmune disease* | *History of autoimmune disease* | *History of autoimmune disease* | *Not implemented (Not in EDM data model)* |
| *History of idiopathic pulmonary fibrosis, organizing pneumonia (e.g., bronchiolitis obliterans), drug-induced pneumonitis, idiopathic pneumonitis, or evidence of active pneumonitis on screening chest CT scan* | *History of idiopathic pulmonary fibrosis, organizing pneumonia (e.g., bronchiolitis obliterans), drug-induced pneumonitis, idiopathic pneumonitis, or evidence of active pneumonitis on screening chest CT scan* | *History of idiopathic pulmonary fibrosis, organizing pneumonia (e.g., bronchiolitis obliterans), drug-induced pneumonitis, idiopathic pneumonitis, or evidence of active pneumonitis on screening chest CT scan* | *Not implemented (Not in EDM data model)* |
| *Positive test for HIV (prior to randomization)* | *Positive test for HIV (prior to randomization)* | *Positive test for HIV (prior to randomization)* | *Implemented (Same)* |
| *Patients with active hepatitis B (chronic or acute; defined as having a positive hepatitis B surface antigen [HBsAg]* | *Patients with active hepatitis B (chronic or acute; defined as having a positive hepatitis B surface antigen [HBsAg]* | *Patients with active hepatitis B (chronic or acute; defined as having a positive hepatitis B surface antigen [HBsAg]* | *Implemented (Same)* |
| *test at screening) or hepatitis C*<br>- *Patients with past hepatitis B virus (HBV) infection or resolved HBV infection (defined as the presence of hepatitis B core antibody and absence of HBsAg) are eligible. HBV DNA test must be performed in these patients prior to randomization.*<br>- *Patients positive for hepatitis C virus (HCV) antibody are* | *test at screening) or hepatitis C*<br>- *Patients with past hepatitis B virus (HBV) infection or resolved HBV infection (defined as the presence of hepatitis B core antibody and absence of HBsAg) are eligible. HBV DNA test must be performed in these patients prior to randomization.*<br>- *Patients positive for hepatitis C virus (HCV) antibody are* | *test at screening) or hepatitis C*<br>- *Patients with past hepatitis B virus (HBV) infection or resolved HBV infection (defined as the presence of hepatitis B core antibody [HBcAb] and absence of HBsAg) were eligible only if they were negative for HBV DNA.*<br>- *Patients positive for hepatitis C virus (HCV) antibody were eligible only if* | |
| *eligible only if polymerase chain reaction is negative for HCV RNA.* | *eligible only if polymerase chain reaction is negative for HCV RNA.* | *PCR was negative for HCV RNA.* | |
| *Active tuberculosis* | *Active tuberculosis* | *Active tuberculosis* | *Not implemented (Not in EDM data model)* |
| *Severe infections within 4 weeks prior to randomization, including but not limited to hospitalization for complications of infection, bacteremia, or severe pneumonia* | *Severe infections within 4 weeks prior to randomization, including, but not limited to, hospitalization for complications of infection, bacteremia, or severe pneumonia* | *Severe infections within 4 weeks prior to randomization, including but not limited to hospitalization for complications of infection, bacteremia, or severe pneumonia* | *Not implemented (Not in EDM data model)* |
| *Received therapeutic oral or IV antibiotics within 2 weeks prior to randomization* | *Received therapeutic oral or IV antibiotics within 2 weeks prior to randomization* | *Received therapeutic oral or IV antibiotics within 2 weeks prior to randomization* | *Implemented (Likely low completeness.  Using* |
| | | | *EDM data model, it is not possible to distinguish between therapeutic and prophylactic antibiotics, leading to potential for misclassification if this criteria were applied.)* |
| *Significant cardiovascular disease, such as New York Heart Association cardiac disease (Class II or greater), myocardial infarction within 3 months prior to randomization, unstable arrhythmias, or unstable angina* | *Significant cardiovascular disease, such as New York Heart Association cardiac disease (Class II or greater), myocardial infarction, or* ***cerebrovascular accident*** *within the 3 months prior to randomization, unstable* | *Significant cardiovascular disease, such as New York Heart Association cardiac disease (Class II or greater), myocardial infarction, or* ***cerebrovascular accident*** *within 3 months prior to randomization, unstable* | *Not implemented (Not in EDM data model)* |
| | *arrhythmias, or unstable angina* | *arrhythmias, or unstable angina* | |
| *Major surgical procedure other than for diagnosis within 28 days prior to randomization or anticipation of need for a major surgical procedure during the course of the study* | *Major surgical procedure other than for diagnosis within 28 days prior to randomization or anticipation of need for a major surgical procedure during the course of the study* | *Major surgical procedure other than for diagnosis within 28 days prior to randomization or anticipation of need for a major surgical procedure during the course of the study* | *Not implemented (Not in EDM data model)* |
| *Prior allogeneic bone marrow transplantation or solid organ transplant* | *Prior allogeneic bone marrow transplantation or solid organ transplant* | *Prior allogeneic bone marrow transplantation or solid organ transplant* | *Not implemented (Not in EDM data model)* |
| *Administration of a live, attenuated vaccine within 4 weeks before randomization or anticipation that such a* | *Administration of a live, attenuated vaccine within 4 weeks before randomization or anticipation that such a* | *Administration of a live attenuated vaccine within 4 weeks before randomization or anticipation that such a* | *Not implemented (Not in EDM data model)* |
| *live attenuated vaccine will be required during the study* | *live attenuated vaccine will be required during the study* | *live attenuated vaccine would be required during the study* | |
| *Any other diseases, metabolic dysfunction, physical examination finding, or clinical laboratory finding giving reasonable suspicion of a disease or condition that contraindicates the use of an investigational drug or that may affect the interpretation of the results or render the patient at high risk from treatment complications* | *Any other diseases, metabolic dysfunction, physical examination finding, or clinical laboratory finding giving reasonable suspicion of a disease or condition that contraindicates the use of an investigational drug or that may affect the interpretation of the results or render the patient at high risk from treatment complications* | *Any other diseases, metabolic dysfunction, physical examination finding, or clinical laboratory finding giving reasonable suspicion of a disease or condition that contraindicated the use of an investigational drug or that may have affected the interpretation of the results or rendered the patient at high risk from treatment complications* | *Not implemented (Not in EDM data model)* |
| *N/A* | *Patients with illnesses or conditions that interfere with their capacity to understand, follow and/or comply with study procedures* | *Illness or condition that may have interfered with a patient's capacity to understand, follow, and/or comply with study procedures* | *Not implemented*<br>*(Not in EDM data model)* |
| | | | |
| ***Exclusion Criteria Related to Medications*** | | | |
| *NA* | *NA* | *Prior treatment with EGFR inhibitors or ALK inhibitors* | *Implemented* |
| *Any approved anti-cancer therapy, including chemotherapy, or hormonal therapy within 3 weeks prior to initiation of study treatment; the* | *Any approved anti-cancer therapy, including hormonal therapy, within 21 days prior to initiation of study treatment; the following exceptions are* | *Any approved anti-cancer therapy, including hormonal therapy within 21 days prior to initiation of study treatment.* | *Implemented*<br>*(Same)* |
| *following exceptions are allowed:*<br>- ***Hormone-replacement therapy or oral contraceptives;***<br>- ***TKIs approved for treatment of NSCLC discontinued 7 days prior to randomization.***<br>*The baseline scan must be obtained after discontinuation of prior TKIs.* | *allowed:*<br>- ***TKIs approved for treatment of NSCLC discontinued 7 days prior to randomization.***<br>*The baseline scan must be obtained after discontinuation of prior TKIs.* | | |
| *Treatment with any other investigational agent or participation in another* | *Treatment with any other investigational agent with therapeutic intent* | *Treatment with any other investigational agent with therapeutic intent* | *Not implemented (Not necessary since patients receiving clinical* |
| *clinical study with therapeutic intent within 28 days prior to randomization* | *within 28 days prior to randomization* | *within 28 days prior to randomization* | *study drug in a prior line will not be eligible.)* |
| *Prior treatment with CD137 agonists or immune checkpoint blockade therapies, anti-PD-1, and anti-PD-L1 therapeutic antibodies* | *Prior treatment with CD137 agonists or immune checkpoint blockade therapies, anti-PD-1, and anti-PD-L1 therapeutic antibodies* | *Prior treatment with CD137 agonists or immune checkpoint blockade therapies, antiPD-1, and antiPD-L1 therapeutic antibodies* | *Implemented (Same)* |
| *Treatment with systemic immunostimulatory agents (including but not limited to IFNs, IL-2) within 4 weeks or five half-lives of the drug, whichever is longer, prior to randomization* | *Treatment with systemic immunostimulatory agents (including, but not limited to, IFNs, IL-2) within 4 weeks or five half-lives of the drug, whichever is longer, prior to randomization* | *Treatment with systemic immunostimulatory agents (including, but not limited to, interferons, interleukin 2) within 4 weeks or 5 half-lives of the drug, whichever was longer,* | *Not implemented (Likely low prevalence)* |
| | | *prior to randomization* | |
| *Treatment with systemic immunosuppressive medications (including but not limited to* ***prednisone****, cyclophosphamide, azathioprine, methotrexate, thalidomide, and antitumor necrosis factor [anti-TNF] agents) within 2 weeks prior to randomization* | *Treatment with systemic immunosuppressive medications (including, but not limited to, corticosteroids, cyclophosphamide, azathioprine, methotrexate, thalidomide, and anti-tumor necrosis factor [anti-TNF] agents) within 2 weeks prior to randomization* | *Treatment with systemic immunosuppressive medications (including, but not limited to, corticosteroids, cyclophosphamide, azathioprine, methotrexate, thalidomide, and anti-tumor necrosis factor [anti-TNF] agents) within 2 weeks prior to randomization* | *Not implemented (This criteria likely cannot be accurately implemented using data captured in EDM data model. Completeness of medications captured in structured orders is likely low, and describing exposure to those medications requires assumptions.)* |
| | | | |
| ***Exclusion Criteria Related to Chemotherapy*** | | | |
| *Known history of severe* | *Known history of severe* | *History of allergic* | *Not implemented* |
| allergic reactions to platinum-containing compounds or mannitol | allergic reactions to platinum-containing compounds or mannitol | reactions to cisplatin, carboplatin, or other platinum-containing compounds | (Not in EDM data model) |
| Known sensitivity to any component of nab-paclitaxel | Known sensitivity to any component of **paclitaxel** or nab-paclitaxel | N/A | Not implemented<br>(Not in EDM data model) |
| N/A | N/A | Patients with hearing impairment (cisplatin) | Not implemented<br>(Not in EDM data model) |
| Grade ≥ 2 peripheral neuropathy as defined by NCI CTCAE v4.0 criteria | Grade ≥ 2 peripheral neuropathy as defined by NCI CTCAE v4.0 (**paclitaxel and nab-paclitaxel**) | Grade  2 peripheral neuropathy as defined by NCI CTCAE v4.0 (**cisplatin**) | Not implemented<br>(Not in EDM data model) |
| Known history of severe | Known history of severe | N/A | Not implemented |
| *hypersensitivity reactions to products containing Cremophor® EL* | *hypersensitivity reactions to products containing Cremophor® EL* | | *(Not in EDM data model)* |